\documentclass[conference]{IEEEtran}
\usepackage{amsmath,amsfonts}
\usepackage{algorithmic}
\usepackage{algorithm}
\usepackage{array}
\usepackage{textcomp}
\usepackage{stfloats}
\usepackage{url}
\usepackage{verbatim}
\usepackage{graphicx}
\usepackage{cite}
\usepackage{bm}
\usepackage{subcaption}
\usepackage{booktabs} 
\usepackage[T1]{fontenc}

\usepackage{amsthm} 
\usepackage{cleveref} 

\usepackage{etoolbox}

\patchcmd{\thebibliography}
{\settowidth}
{\setlength{\itemsep}{0.5\baselineskip} 
	\setlength{\parskip}{0pt}
	\setlength{\parsep}{1.2pt}
	\settowidth}
{}{}

\begin{document}
	
	\title{Deep Learning based Three-stage Solution for ISAC Beamforming Optimization
	}
	
	\author{\IEEEauthorblockN{Qian Gao*, Ruikang Zhong*, and Yuanwei Liu\textsuperscript{\textdagger}}
		\IEEEauthorblockA{*Queen Mary University of London, London, UK} 
		\IEEEauthorblockA{\textsuperscript{\textdagger}The University of Hong Kong, Hong Kong}
		E-mail: \{q.gao, r.zhong\}@qmul.ac.uk, yuanwei@hku.hk}

\maketitle

\begin{abstract}
	
In this paper, a general ISAC system where the base station (BS) communicates with multiple users and performs target detection is considered. Then, a sum communication rate maximization problem is formulated, subjected to the constraints of transmit power and the minimum sensing rates of users. To solve this problem, we develop a framework that leverages deep learning algorithms to provide a three-stage solution for ISAC beamforming. The three-stage beamforming optimization solution includes three modules: 1) an unsupervised learning based feature extraction algorithm is proposed to extract fixed-size latent features while keeping its essential information from the variable channel state information (CSI); 2) a reinforcement learning (RL) based beampattern optimization algorithm is proposed to search the desired beampattern according to the extracted features; 3) a supervised learning based beamforming reconstruction algorithm is proposed to reconstruct the beamforming vector from beampattern given by the RL agent. Simulation results demonstrate that the proposed three-stage solution outperforms the baseline RL algorithm by optimizing the intuitional beampattern rather than beamforming.

\end{abstract}


\section{Introduction}

Integrated sensing and communication (ISAC), which is also known as dual-functional radar-communication or joint radar-communication, is expected to be a key technology in sixth-generation mobile communication (6G) \cite{ISAC6G}. It facilitates the bursting development of plentiful emerging applications, including but not limited to human-computer interaction \cite{hs}, smart home \cite{smarthome} and internet-of-vehicles \cite{V2X}, by providing high-throughput transmission and accurate sensing simultaneously. On the other hand, communication and radar sensing are both evolving towards larger antennas. ISAC systems can not only benefit the spectrum efficiency and implementation cost reduction by resource sharing and hardware reuse but also introduce mutual gains via communication-aided sensing or sensing-aided communication~\cite{scmode}. Although ISAC has the above advantages, extra efforts are required to execute resource allocation and interference management between communication and sensing functionalities. To tackle these issues, beamforming techniques \cite{multiantenna1, multiantenna2, multiantenna3} have been widely studied, as they enable spatial filtering and efficient use of spectral and spatial resources, thereby facilitating the coexistence of dual functionalities.

Beamforming optimization has the potential to improve the performance gain of multiple-input multiple-output (MIMO) systems, which has been proven to be effective in MIMO communication and MIMO radar systems \cite{mimocomrad}. Following this, a number of researchers have also investigated numerous algorithms to support beamforming optimization in ISAC scenarios. These algorithms can generally be divided into conventional convex optimization based solutions and machine learning based solutions. Convex optimization methods tend to be computational complex when dealing with scenarios involving a large number of parameters \cite{convexslow}. In contrast, machine learning based solutions can adapt to non-convex and high-dimensional problems, and it dose not have significant complexity in inference. However, they cannot guarantee the optimality of beamforming and are weak in interpretation. As the number of antennas and users increases, deep learning (DL) have more advantages in ISAC beamforming optimization than convex optimization. 

Existing deep learning-based beamforming optimization approaches typically attempt to directly map the channel state information (CSI) to the beamforming vector. While these methods have demonstrated promising results, they suffer from two key limitations. First, the dimensionality of the optimization variables grows rapidly with the number of antennas and users, leading to increased computational complexity and training difficulty. Second, directly optimizing the beamforming vector lacks interpretability, making it difficult to understand or control the underlying beamforming behavior.

To address these limitations, in this paper, we propose a three-stage beamforming optimization solution that decomposes beamforming optimization into feature extraction, beampattern optimization, and beamforming reconstrcution. By optimizing the beampattern, we are able to control the directionality and interference behavior in a more structured manner, while significantly reducing the learning complexity. The main contributions of this paper can be summarized as follows: 1) We consider a full-duplex (FD) ISAC system with two ULAs acting as a transceiver, and formulate a beamforming optimization problem to maximize the sum communication rate under sensing rate and power constraints. To derive the optimal transmit and receive beamforming vectors more intuitional, we transform the beamforming optimization problem to a beampattern optimization problem by multiplying corresponding array response vectors; 2) we design a three-stage solution to decompose the optimization problem into three sub-modules: unsupervised learning based CSI feature extraction module, RL-based beampattern optimization module, and supervised learning based beamforming reconstruction module. Compared to direct RL approaches, the proposed three-stage framework enhances generalization via feature extraction, improves training stability by optimizing in the beampattern domain, and offers better interpretability through the intermediate beampattern representation.

\section{System Model and Problem Formulation} \label{section:2}

\begin{figure}[t!]
	\centering
	\captionsetup{justification=centering}
	\includegraphics[width=0.4\textwidth]{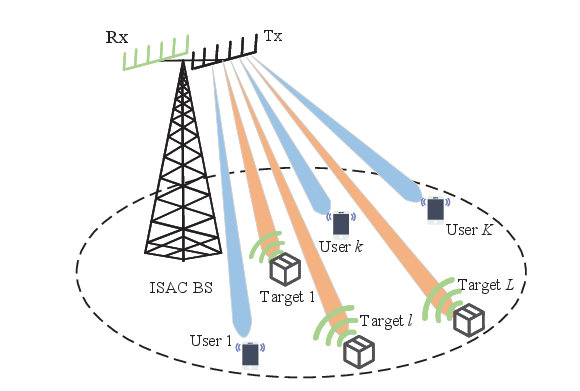}
	\caption{Illustration of the FD ISAC system.}
	\label{Fig.1}
\end{figure}

We consider a FD ISAC system as shown in Fig.~\ref{Fig.1}, where a base station (BS) equipped with $N_t$ transmit antennas, and $N_r$ receive antennas. The transceiver provides downlink communication for $K$ single antenna users and radar sensing for $L$ targets via the same time-frequency resource. To be specific, $N_t$ antennas transmit communication signals to the $K$ users and dedicated sensing signals to $L$ targets simultaneously. Then, $N_r$ antennas receive the echoes from all targets. We assume the BS antenna arrays are ULAs with a half wavelength antenna spacing, $d = \lambda/2$. Thus, the transmit (receive) antenna aperture is $D = (N_t -1)d$. 

\subsection{Signal Model}

We employ a narrowband ISAC signal, $\mathbf{x}\in \mathbb{C}^{N_t \times 1} $, for multi-user downlink communication and multi-target sensing. Following \cite{signalmodel}, the integrated signal can be expressed as
\vspace{-0.3em}
\begin{align}
	\mathbf{x} = \sum_{k=1}^{K}\mathbf{w}_k s_k + \sum_{l=1}^{L}\mathbf{v}_l z_l, \label{1}
\end{align}
where $\mathbf{w}_k \in \mathbb{C}^{N_t \times 1}$ stands for the BS's transmit beamforming vector for user $k$, $k \in \{1, \dots, K\}$, and $s_k \in \mathbb{C}$  is the data symbol for user $k$ with $\mathbb{E}[|s_i|^2]=1$. $\mathbf{v}_l \in \mathbb{C}^{N_t \times 1}$ denotes the transmit beamforming vector associated with target $l$, $l \in \{1, \dots, L\}$, and $z_l \in \mathbb{C}$ with unit power is the dedicated sensing signal for target $l$. Here, the signals $\{s_k\}_{k=1}^{K}$ and $\{z_l\}_{l=1}^{L}$ are assumed to be independent with each other. Moreover, the transmit beamforming vectors should comply with the constraint $\sum_{k=1}^{K}||\mathbf{w}_k||^2 + \sum_{l=1}^{L}||\mathbf{v}_l||^2 \le P_{\text{max}}$, considering transmit ULA's maximum available power budget $P_{\text{max}}$. 

Denoting the the downlink channel between transmit ULA and the $k$-th user as $\mathbf{h}_k \in \mathbb{C}^{N_t \times 1}$, the received signal at user $k$ can be given by
\begin{align}
	y^{\text{com}}_k &= \mathbf{h}_{k}^{H}\mathbf{x} + n_{k} \notag \\ 
	&= \mathbf{h}_{k}^{H}\mathbf{w}_k s_k + \sum_{i=1, i\neq k}^{K}\mathbf{h}_{k}^{H}\mathbf{w}_i s_i + \sum_{l=1}^{L}\mathbf{h}_{k}^{H}\mathbf{v}_l z_l + n_{k} \label{2},
\end{align}
where $n_k \sim \mathcal{CN}(0, \sigma^2_{k})$  is the circularly symmetric complex Gaussian noise with variance $\sigma^2_{k}$.

Meanwhile, the channels between transmit ULA and the $l$-th target and between the $l$-th target and receive ULA are represented as $\mathbf{g}_{l,t} \in \mathbb{C}^{N_t \times1}$ and $\mathbf{g}_{l,r} \in \mathbb{C}^{N_r \times1}$, respectively. The echos reflected from targets is
\vspace{-0.6em}
\begin{align}
	y^{\text{sen}}_{\text{BS}} = \sum_{l=1}^{L}\mathbf{g}_{l,r} \mathbf{g}^{H}_{l,t}\mathbf{x} + \mathbf{H}_{\text{SI}}\mathbf{x} +  \mathbf{n},\label{3}
\end{align}
where $\mathbf{H}_{\text{SI}}\mathbf{x}$ is the self-interference (SI) due to simultaneous transmission and reception, $\mathbf{H}_{\text{SI}} \in \mathbb{C}^{N_r \times N_t}$ is the residual SI channel after SI cancellation \cite{selfinterference}, and $\mathbf{n} \in \mathbb{C}^{N_r \times 1}$ is the circularly symmetric complex Gaussian noise with variance $\sigma^2_{r}$$\mathbf{I}_{N_r}$. According to \cite{radarchannel}, the radar channel can be expressed as the multiplication of complex path-loss coefficient and steering vector. Thus, (\ref{3}) can be rewritten as
\vspace{-0.2em}
\begin{align}
	y^{\text{sen}}_{\text{BS}} = \sum_{l=1}^{L}\alpha_{l}\mathbf{A}(\theta_l)\mathbf{x} + \mathbf{H}_{\text{SI}}\mathbf{x} +  \mathbf{n} , \label{4}
\end{align}
where $\alpha_{l}$ is assumed to be the round path-loss coefficient of target $l$, and $\mathbf{A}(\theta_l) = \mathbf{a}_{r}(\theta_l)\mathbf{a}^{\mathbf{H}}_{t}(\theta_l) \in \mathbb{C}^{N_r \times N_T}$. The transmit steering vector $\mathbf{a}_{t}(\theta_l)$ and receive steering vector $\mathbf{a}_{r}(\theta_l)$ can characterized as
\vspace{-0.5em}

\begin{align}
	\mathbf{a}_{t}(\theta_l) =  \left[1, e^{-j2\pi\Delta_t\sin(\theta_l) }, \dots,e^{-j2\pi(N_t-1)\Delta_t\sin(\theta_l) }\right] , \label{5}
\end{align}
\vspace{-1em}

\begin{align}
	\mathbf{a}_{r}(\theta_l) = \left[1, e^{-j2\pi\Delta_r\sin(\theta_l) }, \dots,e^{-j2\pi(N_r-1)\Delta_r\sin(\theta_l) }\right], \label{6}
\end{align}
where $\Delta_t$ and $\Delta_r$ are normalized intervals between transmit ULA's and receive ULA's adjacent antennas, respectively.

The receive ULA then applies to receive beamforming vectors $\mathbf{u}_l \in \mathbb{C}^{N_r\times 1}$ on the reflected signal $y_{\text{BS}}$ to capture the desired signal from target $l$. Thus, the sensed signal that corresponds to target $l$ is given as

\vspace{-1em}

\begin{align}
	y^{\text{sen}}_l = &\alpha_{l}\mathbf{u}^{H}_l\mathbf{A}(\theta_l)\mathbf{x} \notag\\&+\sum_{j=1, j\neq l}^{L}\alpha_{j}\mathbf{u}^{H}_l\mathbf{A}(\theta_j)\mathbf{x} + \mathbf{u}^{H}_{l}\mathbf{H}_{\text{SI}}\mathbf{x} +  \mathbf{u}^{H}_{l}\mathbf{n}. \label{7}
\end{align}

\subsection{Communication and Sensing Rates}
In this paper, we use communication and sensing rates to evaluate the performance of the FD ISAC system. With (\ref{2}), the communication signal-to-interference-plus-noise ratio (SINR) of $k$-th user can be calculated as

\begin{align}
	\gamma^{\text{com}}_k = \frac{|\mathbf{h}_{k}^{H}\mathbf{w}_k|^2 }{\sum_{i=1, i\neq k}^{K}|\mathbf{h}_{k}^{H}\mathbf{w}_i|^2 + \sum_{l=1}^{L}|\mathbf{h}_{k}^{H}\mathbf{v}_l|^2 + \sigma^2}.  \label{8}
\end{align}
Thereby, the corresponding communication rate is given as
\begin{align}
	\mathcal{R}^{\text{com}}_k = \log (1+\gamma^{\text{com}}_k). \label{9}
\end{align}

Similarly, using (\ref{7}), the sensing SINR of target $l$ is formulated as

\begin{align}
	\gamma^{\text{sen}}_l = \frac{|\alpha_{l}\mathbf{u}^{H}_l\mathbf{A}(\theta_l)\mathbf{x}|^2}{\sum_{j=1, j\neq l}^{L}|\alpha_{j}\mathbf{u}^{H}_l\mathbf{A}(\theta_j)\mathbf{x}|^2 + |\mathbf{u}^{H}_{l}\mathbf{H}_{\text{SI}}\mathbf{x}|^2 +  |\mathbf{u}^{H}_{l}\mathbf{n}|^2}  . \label{10}
\end{align}
Moreover, the sensing rate of target $l$ is obtained as follows:
\begin{align}
	\mathcal{R}^{\text{sen}}_l = \log (1+\gamma^{\text{sen}}_l). \label{11}
\end{align}

Compared with the beampattern gain metric used in \cite{priorwork5}, which only indicates the power assigned towards the target, the sensing rate considers both transmit and receive beampatterns. Sensing rate further feedback on how much information can be obtained from a particular target. Also, the target detection probability is a monotonically increasing function of sensing rate \cite{monotonically}.

\subsection{Problem Formulation} 
In this paper, we aim to maximize the sum communication rate of all users with a constraint transmit power budget while ensuring each target's sensing rate. A joint optimization problem about BS's transmit beamforming vectors $\{\mathbf{w}_k\}_{k=1}^K$, $\{\mathbf{v}_l\}_{l=1}^L$ and receive beamforming vectors $\{\mathbf{u}_l\}_{l=1}^K$ is formulated as follows
\vspace{-1em}
\begin{subequations}
	\begin{align}
		\mathcal{P}1: &\max_{\tilde{\mathcal{A}}}   \sum_{k=1}^{K}  \mathcal{R}^{\text{com}}_k\notag,  \\
		\textrm{s.t.} \ \
		&\sum_{k=1}^{K}\mathbf{w}_k+\sum_{l=1}^{L}\mathbf{v}_l \leq P_{\text{max}},\forall k, \forall l ,\label{OPP4}\\
		&||\mathbf{u}_l||^2 = 1, \forall l , \label{OPP5} \\
		& \mathcal{R}^{\text{sen}}_l \geq \Gamma^{\text{sen}}_{\text{req}},\forall l, \label{OPP6}\\
		& \mathcal{A}=\text{G}^{-1}(\tilde{\mathcal{A}}), \label{OPP7}
	\end{align}
\end{subequations}
where $\tilde{\mathcal{A}} = \left\{\{\text{G}(\mathbf{w}_k)\}_{k=1}^K, \{\text{G}(\mathbf{v}_l)\}_{l=1}^L,
\{\text{G}(\mathbf{u}_l)\}_{l=1}^L\right\}$ is the set of beampattern optimization variables, \eqref{OPP7} ensures that beamforming can be reconstructed from beampattern. The function $\text{G}$ denotes the calculation process from a given beamforming vector to the corresponding beampattern values on different angles. Take the transmit beamforming vector $\mathbf{w}_k$ as an example, its beampattern can be given by
\begin{align}
	\text{G}(\mathbf{w}_k) = \left[\sum_{n=0}^{N-1} \mathbf{w}_{k}[n] \cdot e^{-j 2\pi d n \sin(\theta_{\text{min}})}, \right. & \nonumber \\
	\left. \dots, \sum_{n=0}^{N-1} \mathbf{w}_{k}[n] \cdot e^{-j 2\pi d n \sin(\theta_{\text{max}})} \right] \label{14}
\end{align}
where $\mathbf{w}_{k}[n]$ refers to the $n$-th value of the vector $\mathbf{w}_k$. For the inverse analytical solution $\text{G}^{-1}$, it cannot be found as a weighted sum of multiple complex numbers would form a complex response surface. In this paper, we design a deep learning network to approximate it, and a detailed demonstration will be shown in the next section.

\section{Proposed Solution}

\begin{figure}[t!]
	\centering
	\captionsetup{justification=centering}
	\includegraphics[width=0.5\textwidth]{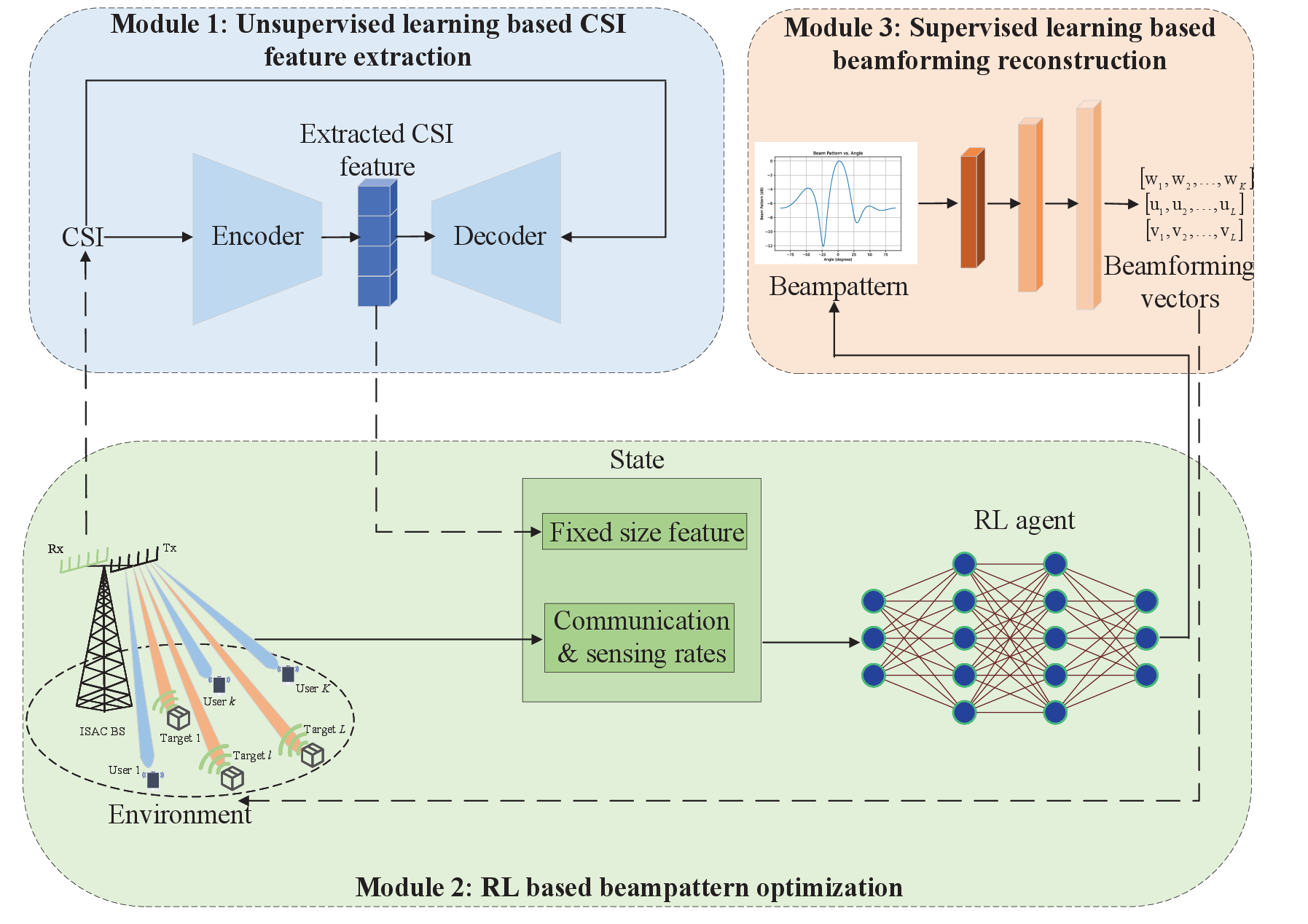}
	\caption{Workflow of the proposed ICT framework.}
	\label{Fig.2}
\end{figure}

\subsection{Unsupervised learning based CSI Feature Extraction}

The first module is an unsupervised learning-based feature extraction module that processes high-dimensional CSI. This is achieved through the Autoencoder (AE) \cite{AE}, which can efficiently learn low-dimensional representations without requiring labelled data. By compressing the high-dimensional CSI data, the module reduces computational complexity.  The details about CSI data generation and the network structure of Autoencoder are illustrated as follows:

\subsubsection{CSI Dataset Generation} In our FD ISAC system, the CSI includes downlink channel $\mathbf{h}_k, \mathbf{g}_{l,r}$, uplink channel $\mathbf{g}_{l,t}$ and the residual SI  channel $\mathbf{H}_{\text{SI}}$. For the given number of transceiver antennas, users and targets, we use set $\mathbf{H} = \{\mathbf{h}_k, \mathbf{g}_{l,r}, \mathbf{g}_{l,t}, \mathbf{H}_{\text{SI}}\}$ to represent a CSI data sample. Then, the locations of users and targets are randomly sampled $D_1$ times within their range of activity to generate the CSI dataset.

\subsubsection{Data Preprocessing} The original CSI data sample $\mathbf{H}$ needs to be preprocessed before feeding into the AE. The process involves transforming a set of complex-valued matrices into a structured real-valued vector representation. First, each matrix $\mathbf{H}_i$ in the set $\mathbf{H}$ is flattened into a column vector $\tilde{\mathbf{h}}_i$ (To distinguish with $\mathbf{h}_k$, we add a tilde on the flattened CSI $\tilde{\mathbf{h}}_i$. ) using the vectorization operation $\text{vec}(\mathbf{H}_i)$, which rearranges all elements of the matrix into a single column
\begin{align}
	\tilde{\mathbf{h}}_i = \text{vec}(\mathbf{H}_i),\quad \forall \mathbf{H}_i \in \mathbf{H}, \label{15}
\end{align}
These flattened vectors are then concatenated to form a single long vector $\mathbf{h}$, representing all matrices in the set 
\begin{align}
	\tilde{\mathbf{h}} = \begin{bmatrix} \tilde{\mathbf{h}}_1,  \tilde{\mathbf{h}}_2, \tilde{\mathbf{h}}_3,  \tilde{\mathbf{h}}_4 \end{bmatrix}^T, \label{16}
\end{align}
Since the matrices contain complex numbers, the final step separates the real and imaginary parts of $\tilde{\mathbf{h}}$, resulting in two independent real-valued vectors. These two vectors are then stacked to construct the final real-valued representation
\begin{align}
	\mathbf{CSI} = \begin{bmatrix} \Re(\tilde{\mathbf{h}}) \\ \Im(\tilde{\mathbf{h}}) \end{bmatrix}, \label{17}
\end{align}
where $\Re(\cdot)$ and $\Im(\cdot)$ are used to indicate the real and imaginary functions, separately. 

\subsubsection{Network Structure} The encoder and decoder of the AE are composed of $k_1$ layers of fully connected neural networks. To be specific, $\mathbf{CSI}$ is first fed into the encoder $E(\cdot, \vartheta_E)$ to obtain the latent feature representation $\mathbf{f}$
\begin{align}
	\mathbf{f} = E(\mathbf{CSI}, \vartheta_E),\label{18}
\end{align}
Then the decoder $D(\cdot, \vartheta_D)$ uses $\mathbf{f}$ to reconstruct $\mathbf{CSI}$:
\begin{align}
	\tilde{\mathbf{CSI}} = D(\mathbf{f}, \vartheta_D),\label{19}
\end{align}
where $\vartheta_E$ and $\vartheta_D$ are parameters of the encoder and decoder, they are updated by minimizing the Mean Square Error (MSE) between the reconstructed $\tilde{\mathbf{CSI}}$ and original $\mathbf{CSI}$
\vspace{-0.5em}
\begin{align}
	MSE = \frac{1}{D}\sum_{i=1}^{D}| \tilde{\mathbf{CSI}}-\mathbf{CSI} |^2\label{20}.
\end{align}

After sufficient training, only the encoder is used to predict extracted CSI features in the testing stage. This feature extraction step ensures that the critical information within CSI is preserved compactly, serving as the foundation for subsequent reinforcement learning based beampattern optimization.

\subsection{RL based Beampattern Optimization}

The second module leverages RL to optimize the beampattern based on the latent feature $\mathbf{f}$ extracted from the CSI and the ISAC performance metrics. By using RL, the module dynamically predicts the optimal beampattern to maximize the system's overall performance, balancing both communication and sensing functionalities.

Unlike traditional beamforming vectors, beampatterns provide a more intuitive representation of power distribution over angle domain. This abstraction simplifies optimization and facilitates the design of robust beampatterns under varying system configurations and channel conditions. To be specific, we uniformly sample $A$ points between -90 degree and 90 degree from the beampattern curve and then use RL to predict the beampattern gain at every point. In this way, the action space size of RL agent is a fixed as $A$ instead of $2 \times (N_t \times (K+L) + N_r \times L)$ in beamforming, which will change when the user or antenna number changes.

In this paper, we use the advantage actor-critic (A2C) algorithm to implement beampattern optimization. A2C is an on-policy, actor-critic reinforcement learning algorithm that combines policy-based and value-based methods. It uses a critic (value function) to reduce variance in policy updates, making training more stable compared to pure policy gradient methods like REINFORCE. Different from A3C (Asynchronous Advantage Actor-Critic), which runs asynchronously, A2C synchronizes updates across multiple environments, leading to more efficient training on modern hardware (especially GPUs). There are two key reasons to use A2C: the first is for sample efficiency, since A2C leverages both policy gradients (actor) and value function (critic), it improves sample efficiency compared to pure policy gradient methods; the other one is due to its compatibility in both discrete and continuous action space. For baseline algorithm\footnote{Baseline algorithm in this paper refers to use unextracted CSI as RL input and then optimize beamforming directly.}, we employ an analog-only beamforming with 3-bit quantized phase shifters considering the low-cost and power consumption in practical deployment. While for the proposed three-stage solution, the beampattern prediction is executed in continuous beampattern gain values.

\subsubsection{State} For baseline algorithm, we define the state $\mathbf{s}_t$ as a vector that consists of CSI $\mathbf{H}_t$ at the time slot $t$ and the sum communication data rate $\mathcal{R}^{\text{com}}_{t-1}$ at time slot $t-1$
\begin{align}
	\mathbf{s}_t = \{\mathbf{H}_t, \mathcal{R}^{\text{com}}_{t-1}\},\\ 
	\mathcal{R}^{\text{com}}_{t-1} = \sum_{k=1}^{K}\mathcal{R}^{\text{com}}_{k,t-1}.
\end{align}
At each time slot $t$, RL agent receives the current CSI and communication metric from last time slot to determine beamforming policy of next time slot. Here, $\mathcal{R}^{\text{com}}_{t-1}$ reflects the beamforming quality at last time slot, which helps agent to improve based on previous experiences. For proposed three-stage solution, we replace CSI with the extracted features $\mathbf{f}$
\begin{align}
	\tilde{\mathbf{s}}_t = \{\mathbf{f}_t, \mathcal{R}^{\text{com}}_{t-1}\}.
\end{align}

\subsubsection{Action} For baseline algorithm, the action $\mathbf{a}_t$ is defined as combination of the real and imaginary part of beamforming vectors
\vspace{-1em}
\begin{align}
	\mathbf{a}_t = \{ \Re(\mathbf{w}), \Re(\mathbf{u}), \Re(\mathbf{v}), \Im(\mathbf{w}), \Im(\mathbf{u}),\Im(\mathbf{v})\}. \label{24}
\end{align}

For proposed three-stage solution, the action $\tilde{\mathbf{a}}_t$ is composed by beampatterns from three beamforming vectors: $\mathbf{w}_k$, $\mathbf{u}_l$ and $\mathbf{v}_l$. For each beamforming vector, we define the following beampattern
\vspace{-1em}
\begin{align}
	p_{\mathbf{w}}(\theta) &= \text{G}(\mathbf{w})= |\mathbf{a}^{\mathbf{H}}_{t}(\theta)\mathbf{w}|^2,\label{25}\\
	p_{\mathbf{v}}(\theta) &= \text{G}(\mathbf{v})= |\mathbf{a}^{\mathbf{H}}_{t}(\theta)\mathbf{v}|^2,\label{26}\\
	p_{\mathbf{u}}(\theta) &= \text{G}(\mathbf{u})= |\mathbf{u}^{\mathbf{H}} \mathbf{a}_{r}(\theta)|^2.\label{27}
\end{align}

The RL agent outputs $A$ samples for each beamforming vector to predict its beampattern 
\vspace{-0.5em}
\begin{align}
	\tilde{\mathbf{a}}_t = \{A_\mathbf{w}, A_\mathbf{v}, A_\mathbf{u}\},
\end{align}
where $A_\mathbf{w}, A_\mathbf{v}, A_\mathbf{u}$ are samples corresponding to $p_{\mathbf{w}}(\theta)$, $p_{\mathbf{v}}(\theta)$ and $p_{\mathbf{u}}(\theta)$, respectively.
\subsubsection{Reward Function}The reward function is consistent for all settings. It is defined as the function of communication and sensing rates. Due to the optimization problem is to maximize the sum communication data rate under the constraints of sensing rates larger than $\Gamma^{sen}_{\text{req}}$, we define the reward function as
\begin{align}
	r& =\begin{cases}
		\sum_{k=1}^{K}\mathcal{R}^{\text{com}}_k &\text{if }  \mathcal{R}^{\text{sen}}_l \ge \Gamma^{\text{sen}}_{\text{req}}, \mathcal{R}^{\text{com}}_k \ge \Gamma^{\text{com}}_{\text{req}}, \forall l, k, \\
		0 &\text{if }  \mathcal{R}^{\text{sen}}_l \ge \Gamma^{\text{sen}}_{\text{req}}, \mathcal{R}^{\text{com}}_k < \Gamma^{\text{com}}_{\text{req}}, \forall l, k, \\
		-1 &\text{if }  \mathcal{R}^{\text{sen}}_l < \Gamma^{\text{sen}}_{\text{req}}, \mathcal{R}^{\text{com}}_k < \Gamma^{\text{com}}_{\text{req}}, \forall l, k.
	\end{cases}
\end{align}

This piecewise function considers three situations: the reward is the sum of the communication rates when both communication and sensing rates meet the minimum requirements; the reward is 0 when the minimum sensing rates are satisfied, but the minimum communication rates are not; and the reward is -1 when either communication rate or sensing rate are not reached. The additional requirement of a minimum communication rate is added here to protect the fairness of communication users.

\subsubsection{Network Structure} 
The A2C algorithm consists of two neural networks: the actor network and the critic network. Actor network $\pi_{\Theta}(\tilde{\mathbf{a}}|\tilde{\mathbf{s}})$ is responsible for outputting the probability distribution over actions (for discrete action spaces) or a parameterized policy (for continuous action spaces).
Critic network $V_{\phi}(\tilde{\mathbf{s}})$ estimates the value function $V(s)$, which helps reduce variance in policy updates. Each network has a single $k_2$ hidden layer and uses ReLU activation.

Through continuous interaction with the environment, the RL agent learns to explore and exploit the action space, ensuring that the optimized beampattern meets the system's requirements for both communication and sensing. This module provides an intuitive and adaptable way to design beampatterns for complex scenarios

\subsection{Supervised learning based Beamforming Reconstruction}

The third module is responsible for reconstructing the beamforming vectors for specific antennas and users based on the optimized beampattern predicted by the reinforcement learning module. This is achieved using a supervised learning approach, where the module maps the beampattern to the corresponding beamforming vectors through a pre-trained DNN network.

The key steps of the reconstruction include:
\subsubsection{Reconstruction Dataset Generation}
For given numbers of antennas and users, we first randomly generate $D_2$ real and imaginary parts of the beamforming vectors and then combine them as in (\ref{24}). The obtained beamforming vectors are then normalized to ensure that their energy complies with the system constraints (\ref{OPP4}) and (\ref{OPP5}). Next, we compute the corresponding beampatterns according to (\ref{25}), (\ref{26}), and (\ref{27}). These beampatterns are then sampled at $A$ discrete angles, allowing us to create discrete representations that align with beamforming vectors. This process ensures that each beamforming vector is uniquely associated with a corresponding discrete beampattern, forming the basis of our dataset.

\subsubsection{Network Structure} We simply use a fully connected neural network $\text{G}^{-1}(\cdot,\vartheta_G)$ with $k_3$ hidden layer to implement the beamforming reconstruction. Batch normalization and ReLU activation are employed for stability and non-linearity. 

\subsubsection{Data Postprocessing} Similar to the data preprocessing in module 1, the output structured real-valued beamforming vector needs to be transformed into complex-valued beamforming vectors for practical implementation.

By decoupling the optimization of beampatterns from the generation of beamforming vectors, this module enables a modular design that enhances system flexibility. Furthermore, the supervised learning approach ensures high accuracy and computational efficiency during deployment, making it suitable for real-time ISAC applications. In the testing stage, the pre-trained reconstruction model can be connected directly after the RL model to recover the beamforming vectors from the optimal beampattern.

\section{Numerical Results}\label{section:4}

In this section, experimental analysis are provided to vadilate the effectiveness of the proposed three-stage solution.
The considered FD ISAC scenario has been shown in Fig. \ref{Fig.1}. To be specific, both the transmit and receive ULAs have $N_t$ = $N_r$ = 4 antennas distributed uniformly at half of the wavelength, i.e., 5.77 cm under the carrier frequency of 2.6 GHz. There are $K = 2$ communication users and $L = 1$ sensing target. For simplicity, we adapt the LoS channel model for all communication users, where $\mathbf{h}_k = \sqrt{\xi_k}\sqrt{\mathbf{N}_t}\mathbf{a}_t(\theta_k)$, $\forall k$, where $\xi_k$ denotes the path loss. The maximum transmitting power of the UPA $P_{\text{max}}$ and the noise power $\sigma^2$ are set to 30 dBW and -60 dBm, respectively. The round path loss coefficient $\alpha_l$ is set to -60 dB. For the residual SI channel $\mathbf{H}_{\text{SI}}$, we follow \cite{HSI} and fix it at -110 dB. The required rate thresholds for communication and sensing are set to $\Gamma_{\text{req}}^{\text{com}}$ = 0.3 bps/Hz and $\Gamma_{\text{req}}^{\text{sen}}$ = 5 bps/Hz. For the proposed three-stage solution, the feature size for extracted CSI is 16 and the sample points for each beampattern is 180.

To investigate the function of each module in the three-stage solution, we define an ablation experiment with 4 cases of the three-stage solution and compare them with the baseline algorithm. \textbf{Case 1} remains modules 1 and 2 and removes module 3; this case helps us know how much information will be lost during the CSI feature extraction and how this loss will influence the ISAC performance. Inversely, \textbf{Case 2} removes module 1 and remains module 2, 3. This case is used to verify the assumption that the beampattern is more intuitional for the RL agent to optimize than the beamforming vector. The \textbf{Case 3} is the original three-stage solution without any modifications; it serves as a reference.  For consistency, we mark the baseline algorithm as \textbf{Case 0}. 
The same network settings and hyperparameters are utilized across all cases. For example, the episode length is 100 steps, and we run 500 episodes for each case. The learning rates of actor and policy networks are $1\times10^{-4}$. The activation function is ReLU, and the optimizer is Adam.

\begin{figure*}
	\centering
	\captionsetup{justification=centering}  
	
	\begin{subfigure}{0.3\textwidth}
		
		\includegraphics[width=\linewidth]{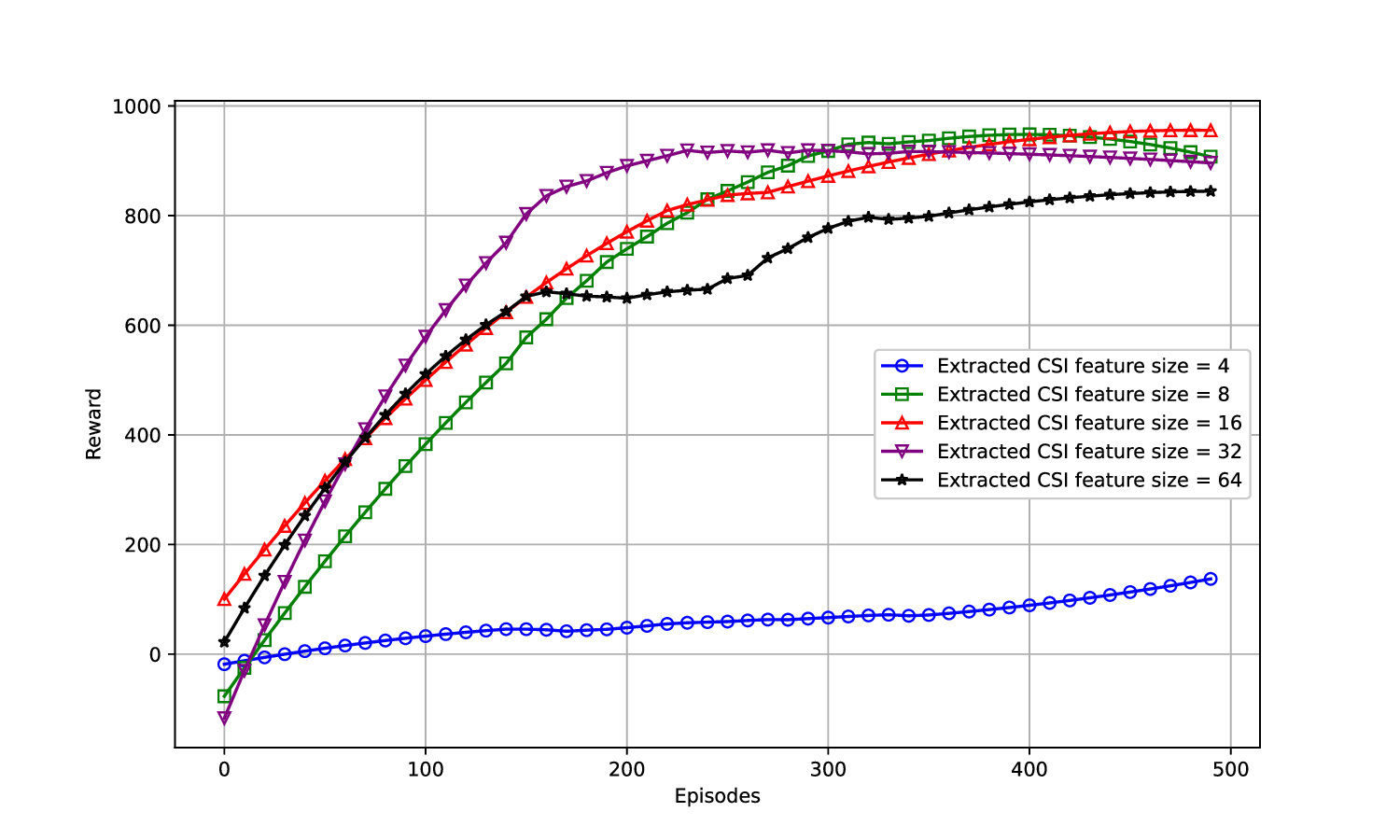}
		\caption{\parbox[c]{4.2cm}{\centering Extracted CSI feature size selection.}}
		\label{Fig.6(a)}
	\end{subfigure}
	\hfill
	\begin{subfigure}{0.3\textwidth}
		
		\includegraphics[width=\linewidth]{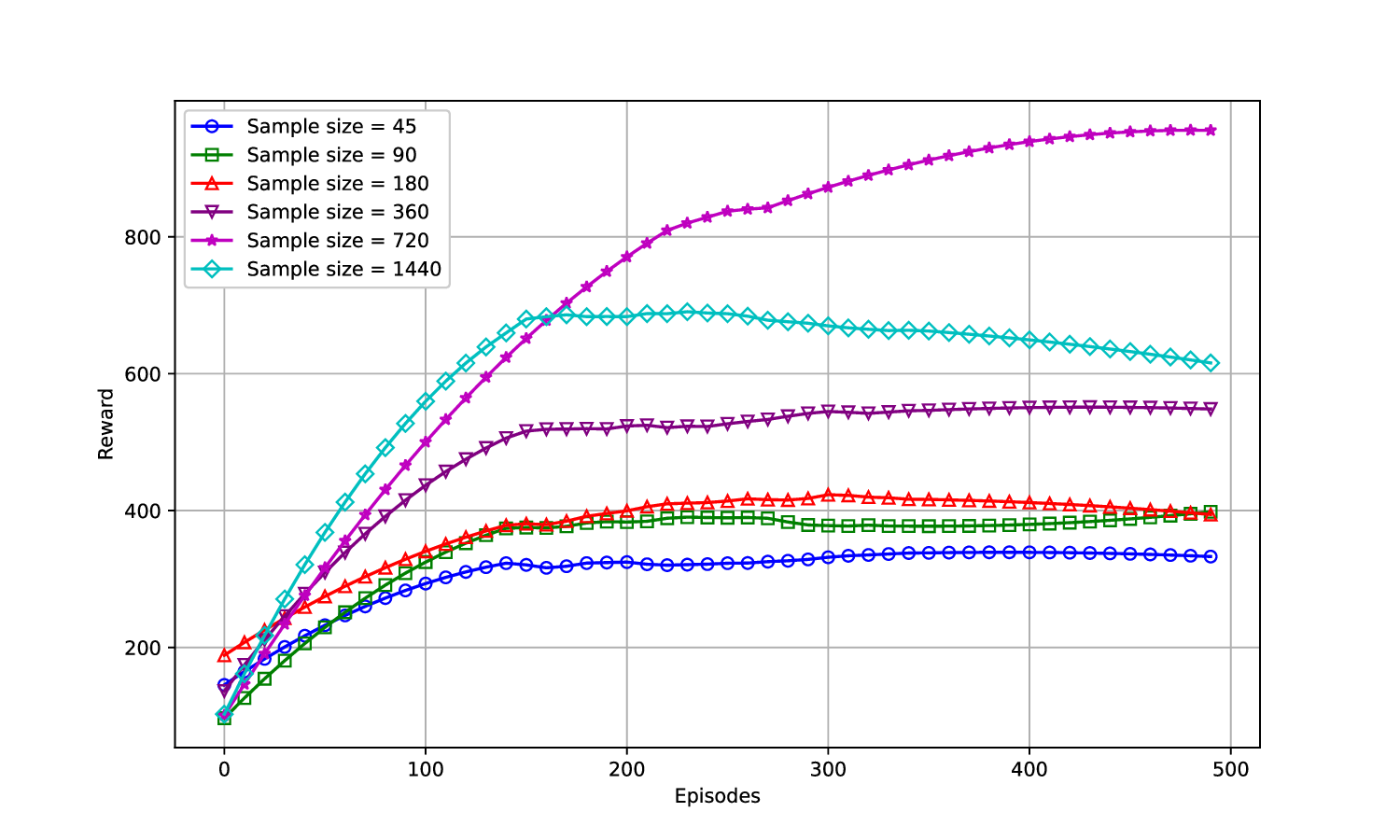}
		\caption{\parbox[c]{4.2cm}{\centering Beampattern sample size selection.         }}
		\label{Fig.6(b)}
	\end{subfigure}
	\hfill
	\begin{subfigure}{0.3\textwidth}
		
		\includegraphics[width=\linewidth]{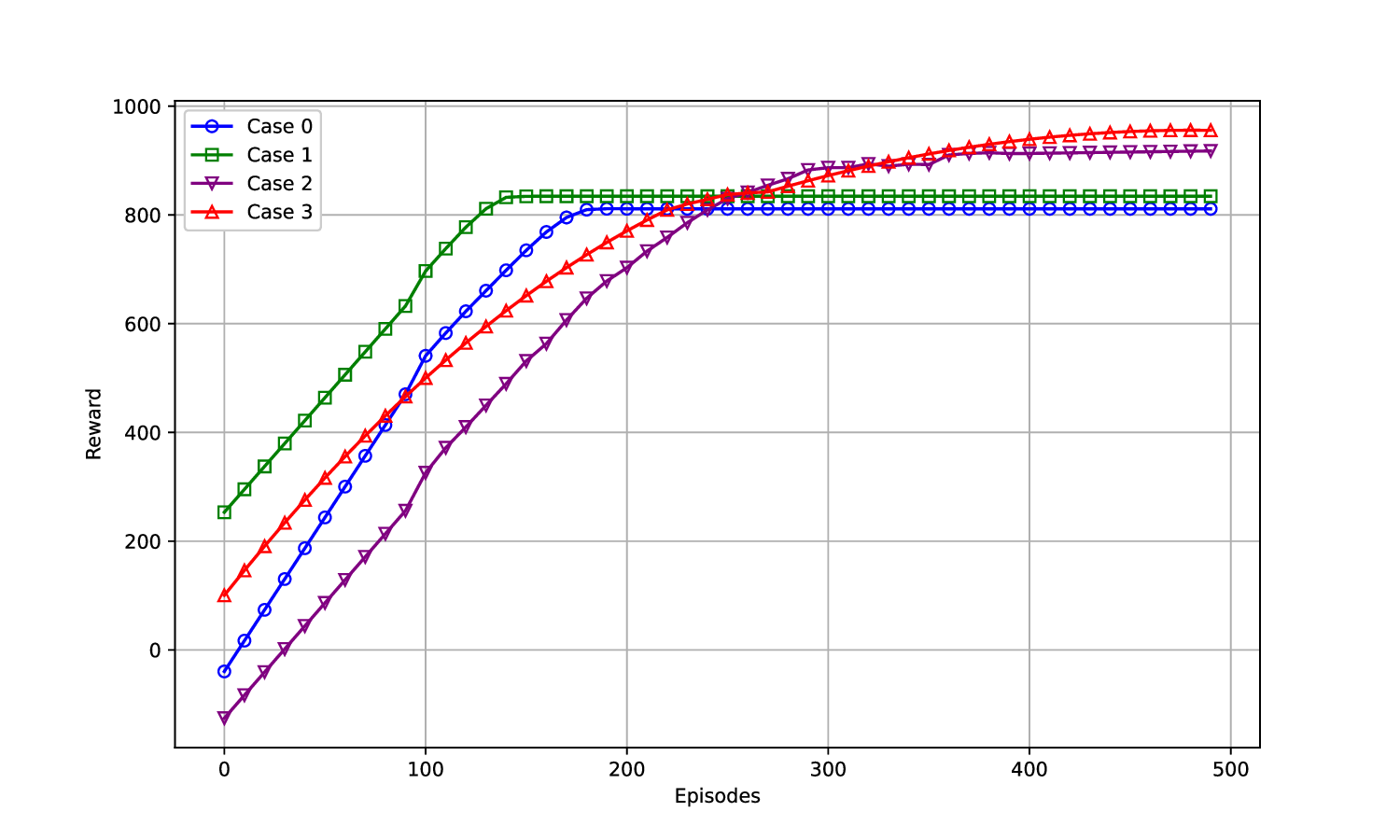}
		\caption{\parbox[c]{4.2cm}{\centering Performance comparison between different cases.}}
		\label{Fig.6(c)}
	\end{subfigure}
	\hfill
	\caption{Hyperparameter selection and performance comparison.}  
	\label{Fig.6}
\end{figure*}

Fig.~\ref{Fig.6(a)} examines the effect of CSI feature size. A too-small size (e.g., \textbf{4}) leads to information loss and poor performance. Sizes \textbf{8} and \textbf{16} significantly enhance learning, with \textbf{16} offering the best trade-off between compression and performance. Size \textbf{32} shows slight further improvement, though with slower convergence. Size \textbf{64} accelerates early learning but results in a lower final reward, likely due to overfitting. Overall, feature sizes of \textbf{16} and \textbf{32} are most effective.

Fig.~\ref{Fig.6(b)} analyzes the effect of beampattern sample size. Smaller sizes (\textbf{45}, \textbf{90}, \textbf{180}) enable fast convergence but low rewards (\textless 400), indicating insufficient spatial resolution. A moderate size \textbf{360} improves both convergence and final reward (\textgreater 500), striking a good balance. The highest performance occurs at \textbf{720}, demonstrating the benefit of refined spatial representation. However, \textbf{1440} degrades final performance despite early gains, due to optimization difficulty in a larger state space.

Fig.~\ref{Fig.6(c)} compares the training reward curves of the baseline \textbf{Case 0} and the proposed three-stage solution under three configurations. \textbf{Case 0} converges around 170 episodes with a stable reward of ~800. \textbf{Case 1}, incorporating Module 1 (CSI compression), converges faster (around 130 episodes) with comparable performance, showing improved training efficiency. \textbf{Case 2}, using only Module 3 (beampattern-based output), learns more slowly but eventually surpasses \textbf{Case 0} with a reward above 850, demonstrating the benefit of structured output. \textbf{Case 3}, combining both modules, achieves the best performance (reward \textgreater 900), balancing efficient learning and high policy quality.
\vspace{-0.5em}
\section{Conclusion}\label{section:5}
In this paper, we proposed a three-stage solution to optimize the beamforming in a interpretable way. Specifically, beamforming optimization was formulated to maximize the sum communication data rate while satisfying the sensing data rate threshold and power constraint. The three-stage solution decompose the beamforming optimization into three sub-questions and solves them with three sub-modules: an AE was used to extract the CSI feature, and an RL agent was used to find the optimal beampattern. Finally, a DNN network reconstructed the desirable beamforming from the beampattern. Simulations showed that the proposed three-stage solution outperforms baseline algorithm through optimizing beampattern and reconstructing beamforming.

\bibliographystyle{IEEEtran}
\bibliography{ref}

\end{document}